\documentclass[twocolumn,pre,superscriptaddress,aps,nofootinbib,10pt]{revtex4-2}

\usepackage{epsfig}
\usepackage{amsmath}
\usepackage{graphicx}
\usepackage{amsmath}
\usepackage{amsthm}
\usepackage{multirow}
\usepackage{relsize}
\usepackage{amssymb}
\usepackage{bm}
\usepackage{textcomp}
\usepackage{epstopdf}
\usepackage{soul}
\RequirePackage{color}

\usepackage{graphicx}
\usepackage{dcolumn}

\usepackage[hidelinks,colorlinks=true,linkcolor=blue,citecolor=blue,urlcolor=blue]{hyperref}

 \usepackage[normalem]{ulem}

\DeclareMathAlphabet{\mathitbf}{T1}{cmr}{bx}{it} 

\begin{document}
\title{Emergence of the 3D diluted Ising model universality class in a  mixture of two magnets
}

\author{J. J. Ruiz-Lorenzo}
\email{ruiz@unex.es}
\affiliation{Departamento de F\'{\i}sica
  and Instituto de Computaci\'{o}n Cient\'{\i}fica Avanzada (ICCAEx),
  Universidad de Extremadura, 06071 Badajoz, Spain}

\author{M. Dudka}
\email{maxdudka@icmp.lviv.ua}
\affiliation{Institute for Condensed Matter Physics, National Academy of Sciences of Ukraine, 79011 Lviv, Ukraine}
\affiliation{${\mathbb L}^4$ Collaboration \& Doctoral College for the Statistical Physics of Complex Systems, Leipzig-Lorraine-Lviv-Coventry, Europe} 
\affiliation{Lviv Polytechnic National University, 79013 Lviv, Ukraine} 

\author{M. Krasnytska}
\email{kras$\_$mariana@icmp.lviv.ua}
\affiliation{Institute for Condensed Matter Physics, National Academy of Sciences of Ukraine, 79011 Lviv, Ukraine}
\affiliation{${\mathbb L}^4$ Collaboration \& Doctoral College for the Statistical Physics of Complex Systems, Leipzig-Lorraine-Lviv-Coventry, Europe} 
\affiliation{ Haiqu, Inc., Shevchenka St, 120G, 79039 Lviv, Ukraine} 

\author{Yu. Holovatch}
\email{hol@icmp.lviv.ua}
\affiliation{Institute for Condensed Matter Physics, National Academy of Sciences of Ukraine, 79011 Lviv, Ukraine}
\affiliation{${\mathbb L}^4$ Collaboration \& Doctoral College for the Statistical Physics of Complex Systems, Leipzig-Lorraine-Lviv-Coventry, Europe} 
\affiliation{Centre for Fluid and Complex Systems, Coventry University, Coventry, CV1 5FB, United Kingdom}
\affiliation{Complexity Science Hub Vienna, 1030 Vienna, Austria}

\date{November 25, 2024}

\begin{abstract} Usually, the impact of structural disorder on the magnetic phase transition in the 3D Ising model is analyzed within the framework of quenched dilution by a non-magnetic component, where some lattice sites are occupied by Ising spins, while others are non-magnetic. This kind of quenched dilution, according to the Harris criterion, leads to a change in the critical exponents that govern the asymptotics in the vicinity of the phase transition point. However, the inherent reason for the emergence of a new, random Ising model universality class is not the presence of a non-magnetic component but the disorder in structure of spin arrangement. To demonstrate this fact, in this paper, we set up extensive Monte Carlo simulations of a random mixture of two Ising-like magnets that differ in spin length $s$ and concentration $c$. In doing so, we analyze the effect of structural disorder \textit{per se} without appealing to the presence of a non-magnetic component. We support our numerical simulations with renormalization group calculations. Our results demonstrate the emergence of the 3D randomly diluted Ising model  universality class in a random mixture of two Ising magnets. While the asymptotic critical exponents coincide with those known for the site-diluted 3D Ising model, the effective critical behavior is triggered by parameters $s$ and $c$. The impact of their interplay is a subject of detailed analysis.
 
\end{abstract}

\maketitle

\section{Introduction} \label{intro}

The quenched structural disorder is inevitably present in real magnetic materials, inducing interesting features of their thermodynamic properties discriminating them from ideally structured counterparts. Therefore disordered magnets became widely used to analyze cooperative phenomena in systems with quenched disorder, for instance, in the vicinity of their phase transitions \cite{Fisher88}.  In particular, studies of the effects of disorder on critical behavior are of great importance not only from a fundamental point of view but also due to novel technological applications of structurally disordered materials \cite{Order_books}.

The consequences of disorder depend on its nature \cite{Holovatch02}. In this paper, we address the case in which random structural imperfections are coupled to the local energy density (or, equivalently, a quenched local random transition temperature).  
The Harris criterion gives a qualitative answer about possible changes in the universality class under the influence of such weak quenched disorder 
\cite{Harris74}. It relates the change in the universality class of the structurally disordered system to the behavior of the heat capacity of its regular (pure) counterpart.
If the heat capacity of the regular system diverges, i.e. if the heat capacity critical exponent $\alpha_{\rm reg} > 0$, then the structural disorder suppresses such divergency leading to the new asymptotics of the heat capacity with $\alpha_{\rm dis} < 0$. It induces also the changes in other critical exponents, amplitude ratios and scaling functions, leading to 
the emergence of a new universality class. Since out of most common $O(m)$-symmetrical magnets only the Ising magnets ($m=1$) are characterized by the diverging heat capacity at space dimension $d=3$, the main focus in the analysis of the emergence of the new universality class has been on the behavior of structurally-disordered quenched Ising magnets.

An archetypal example is given by diluted uniaxial
magnetic  alloys Fe$_x$Zn$_{1-x}$F$_{2}$, Mn$_x$Zn$_{1-x}$F$_2$  prepared by a substitution of a non-magnetic isomorph ZnF$_2$ for its antiferromagnetic counterpart (FeF$_2$ or MnF$_2$) \cite{Birgenau83,Belanger86,Mitchell86}.  Although it is known that ``random magnets are substitutionally disorder materials in which several kinds of magnetic or non-magnetic ions are alloyed together"  \cite{Fisher88}, still the most common examples of magnetic systems with local random transition temperature studied in experiments and  Monte Carlo simulations are represented by diluted systems of magnetic and non-magnetic components.
We refer to, for example,  Refs.~\cite{Ballesteros98,Calabrese03,Ivaneyko05,Hasenbusch07} for the numerical studies of these materials based on the 3D site-diluted Ising model. 

\begin{table*}[htb!]
\caption{Universal exponents and cumulants for three-dimensional
Ising models. As usual, $\nu$ is the correlation length exponent, $\eta$ is the anomalous dimension of the field, $\omega$ is the correction-to-scaling exponent, 
$R_\xi=\xi/L$, $U_4$ is the Binder cumulant and $g_2$ is a cumulant measuring the lack of self-averageness of the systems (the last three cumulants computed at the critical point). 
First two rows: analytic estimates for the diluted Ising model
universality class based on the resumed six-loop renormalization-group functions obtained within 
massive $d=3$ approach (RG, 3D) \cite{Pelissetto00} and minimal subtraction scheme (RG, MS) \cite{Kompaniets21}. 
Number in {\it italic} gives  $\eta$ calculated from the scaling relation $\eta=2-\gamma/\nu$.
The third and fourth rows: numerical Monte-Carlo (MC) simulations results for the exponents and 
cumulants of the 3D site-diluted Ising model. We provide the critical exponents for the 3D bond-diluted Ising model in the fifth column (which belongs to the same universality class as the site-diluted one). The sixth row: our MC results for the mixture of two Ising magnets
differing in spin length, see Table \ref{table:extraQA} and Sec. \ref{MCresults} for more details. In the four and sixth rows we provide the values of the first two leading correction-to-scaling exponents (hereafter, denoted as  $\omega_1$ and $\omega_2$). The 
last row shows numerical estimates for the  regular Ising model.}  
\centering
\label{table:data}
\begin{tabular}{|c|c|c|c|c|c|c|}
\hline
Model & $\nu$ & $\eta$ & $\omega$ & $R_\xi$ & $U_4$ & $g_2$\\
\hline\hline
RG, 3D~\cite{Pelissetto00} & 0.675(19)     & {\it 0.024(79)}    & 0.15(10)& - & - & -  \\
RG, MS~\cite{Kompaniets21} & 0.678(10)    & 0.030(3)      & 0.25(10)& - & - & -  \\
MC, Ising (site-diluted)~\cite{Ballesteros98}        &  0.684(5)   & 0.037(4)      & 0.37(6)        & 0.598(4) & 0.449(6)   & 0.145(3)\\ 
MC, Ising (site-diluted)~\cite{Hasenbusch07}        &  0.683(2)     & 0.036(2)      & 0.33(3) and $0.82(8)$  & - &  -   &-  \\  
MC, Ising (bond-diluted)~\cite{Fytas:10}        &  0.684(7)   & 0.033(3)      &   -   & - &  - & -\\ 
Mixture of two Ising magnets, this paper                                 & 0.678(2)    & 0.033(7)      & 0.31(12) and 0.94(15) & 0.5990(8) & 0.453(3)  & 0.138(5)  \\
MC, Ising~\cite{ferrenberg:18,simmons:17}     &  0.629912(86) & 0.0362978(20) & 0.8303(18)     &0.6431(1)          & 0.46548(5) & 0 \\ \hline \hline
\end{tabular}
\end{table*}

From the theoretical perspective, the renormalization-group (RG) approach relates the emergence of the new universality class to the presence of the new random Ising fixed point of the RG transformation \cite{Pelissetto02,Folk03}. Currently, the high-order RG series have been obtained and carefully analyzed to quantify critical behavior in the new universality class. Since very often the changes of critical properties of structurally disordered magnets are discussed in the context of presence of non-magnetic impurities, the new  universality class is often named the 3D diluted Ising model universality class. Typical numerical values of the universal critical exponents and
cumulants are given in Table \ref{table:data}. They can be compared with the corresponding values of the regular  3D Ising model shown in the last row of the Table. \ref{table:data}.

In our previous paper \cite{Dudka23},  considering the model for the random mixture of different magnetic compounds, we have shown that impurities are non necessary to be non-magnetic to cause the same effects on the critical behavior and the only key ingredient for its modifications is structural disorder itself. To this purpose, we appealed to the model of Ising-like spins of different lengths 
 that was firstly introduced in the context of complex networks \cite{Krasnytska20,Krasnytska21,Krasnytska24} and adopted it for the regular lattice case. Using the replica method we have shown \cite{Dudka23} that the Ising model with randomly distributed elementary spins of different lengths  and the site-diluted Ising model (which contains magnetic and non-magnetic sites) are both described by the field-theoretical $\varphi^4$   effective Hamiltonians of the same symmetry, sharing therefore, the same universality class \cite{Grinstein76,Pelissetto00, Pelissetto02,Folk03,Kompaniets21}.  We have also analyzed the effective critical behavior of the mixture of two Ising magnets differing in spin length.
 
 A confirmation of our findings was obtained for the two-dimensional Ising model with 
 spins of two lengths \cite{Miranda24} where critical exponents of pure two-dimensional Ising model were observed, that according to the Harris criterion \cite{Harris74} is expected also for the two-dimensional Ising random-site model.

The goal of this paper is to provide numerical support for the theoretically predicted scenario of the mere influence of structural disorder on the magnetic phase transition without resorting to dilution by the non-magnetic component. To this end, we set up intensive Monte Carlo 
simulations of the random binary mixture of Ising magnets with spin length ratio $s$ and concentration $c$ on a three-dimensional lattice. 
  Our main result is summarized in the fifth row of Table 
\ref{table:data}, where we give the values of the critical exponents and universal cumulants for the 3D mixture of the Ising magnet with smallest corrections to scaling ($s=1.7$ and  $c=0.53$). To the best of our knowledge, this is the first numerical simulation that convincingly shows the emergence of the 3D diluted Ising universality class for the model with quenched disorder without resorting to the presence of non-magnetic components.

In addition, we will test, at a qualitatively level,  the predictions of the perturbative RG on the behavior of the effective critical exponents confronting them with the results from numerical simulations.

The rest of this paper is organized as follows. In Sec. \ref{modelnalytical}, we introduce the model 
of a random mixture of two Ising-like magnets. Such a model enables one to analyze the effect of structural disorder \textit{per se} without appealing to the presence of a non-magnetic component.
We discuss some results for the asymptotic critical behavior of such a model and show that it belongs to the 3D diluted Ising model universality class. We also show that,
depending on the particular choice of $s$ and $c$, the critical behavior may be influenced by strong corrections to scaling, hence, the effective critical exponents
may be observed.  We 
describe our extensive Monte Carlo simulations in Sec. \ref{MCsimulations}. Our results demonstrate the emergence of the random Ising model universality class in a random mixture of two Ising magnets. While the asymptotic critical exponents coincide with those known for the 3D site-diluted Ising model, the effective critical behavior is triggered by parameters $s$ and $c$.
 Finally, in Sec. \ref{discussion}, we present our results and discuss their implications. Supplementary calculations and discussions are provided in Appendices.

\section{Model and analytical results} \label{modelnalytical}

In this section we describe the model that will be used in Sec. \ref{MCsimulations} for numerical simulations.
We also summarize some results from its renormalization group analysis \cite{Dudka23} and suggest possible scenarios
for the effective critical behavior.

\subsection{Random spin length Ising model}\label{model2lengths}

\begin{figure}[h!]
    \begin{center}
\includegraphics[width=0.17\paperwidth]{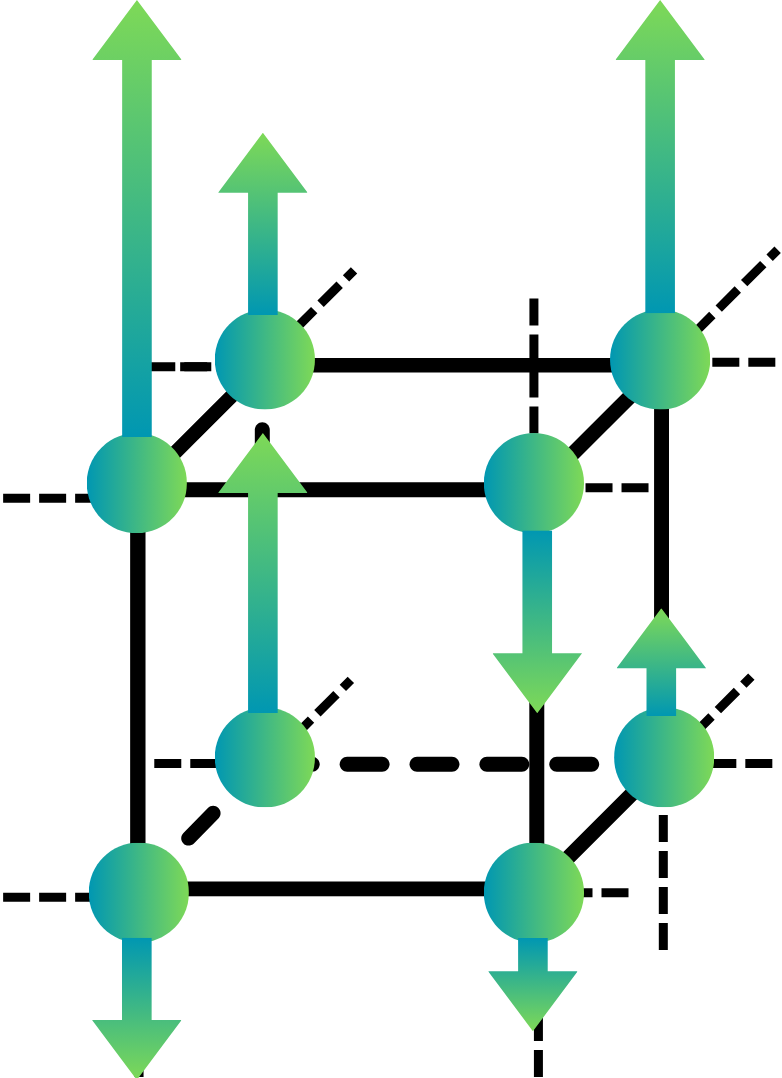}
  \includegraphics[width=0.17\paperwidth]{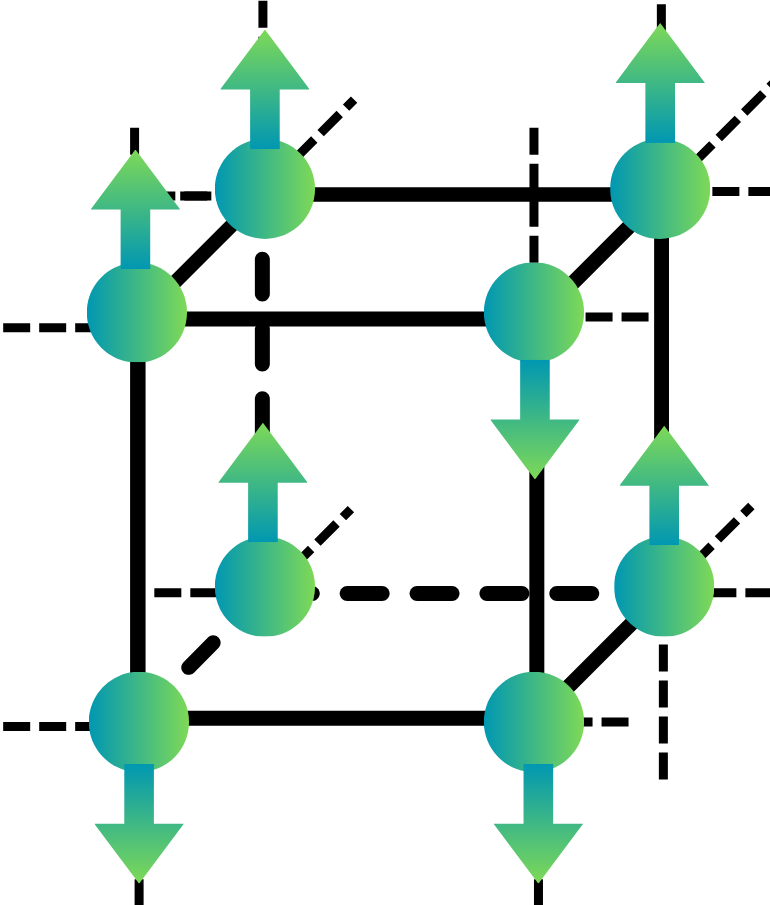}   
  
   {    {\bf (a)} \hspace{3cm} {\bf (b)}}
         
        \includegraphics[width=0.17\paperwidth]{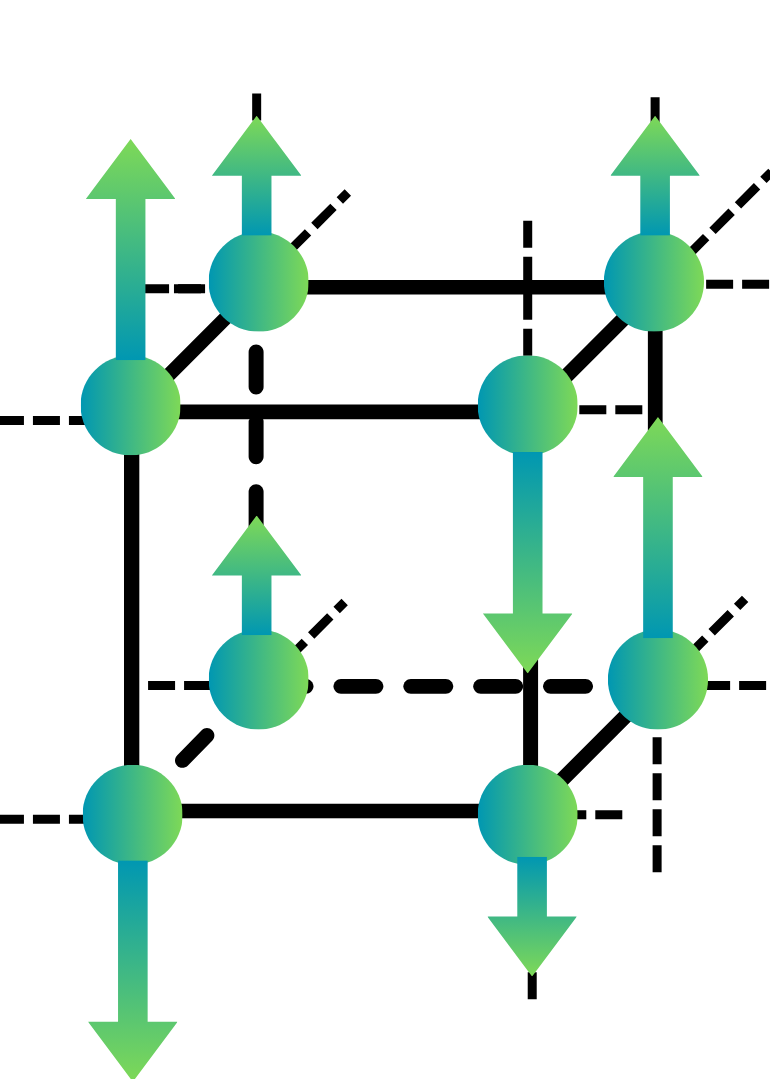}
          \includegraphics[width=0.17\paperwidth]{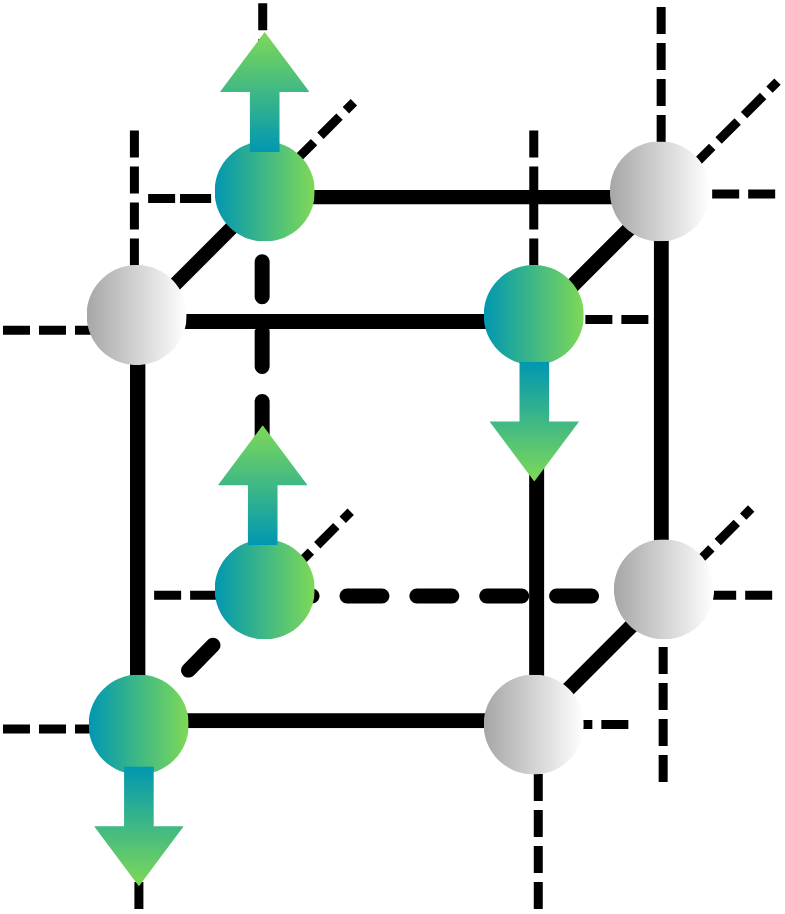}    
          
         {\bf (c)} \hspace{3cm} {\bf (d)}
         
        \caption{Schematic representation of the {\bf (a)} 
        random spin length Ising model, when the lengths of spins are i.i.d.
        random variables, {\bf (b)} standard 3D Ising model (with spins $\pm 1$), 
        {\bf (c)} particular case of the random spin length Ising model: mixture of 
        two Ising magnets with two different spin lengths (that will be analyzed henceforth),
        {\bf (d)} diluted Ising model, where some sites (grey) are occupied by non-magnetic 
        compounds or are empty.  \label{3dIsing}}
    \end{center}
\end{figure}

In this paper we consider a lattice model describing a system of interacting classical
elementary magnets, `spins'. A notorious distinction of this model from a classical Ising model is
that although the spins keep Ising model symmetry (pointing only up and down) they can vary in length.
The Hamiltonian of the random spin length Ising model in the absence of an external magnetic field 
reads \cite{Krasnytska20}:
\begin{equation}
{\cal H} =  - \frac{1}{2}\sum_{{\mathitbf r} \neq {\mathitbf r}^\prime} J(|{\mathitbf r}-{\mathitbf r}^\prime|)
S_{\mathitbf r} S_{{\mathitbf r}^\prime}\, , \hspace{1em}  S_{\mathitbf r} = \pm \hat{L}_{\mathitbf r} .
\label{11.1}
\end{equation}
where the sum spans the sites of a $d$-dimensional hypercubic lattice, $J(r)$ is a short-range interaction 
and $\hat{L}_{\mathitbf r}$ are i.i.d. quenched random variables governed by the distribution function 
\begin{equation}
P(\{\hat{L}\})=\prod_{\mathitbf r} p(\hat{L}_{\mathitbf r}).
\label{11.2}
\end{equation}
Originally, the Hamiltonian (\ref{11.1}) has been suggested in the
context of  collective behavior in complex systems \cite{Krasnytska20,Krasnytska21} and then it was
further applied to analyze magnetic and spin-glass phase transitions in random magnets \cite{Dudka23,Miranda24}. 
A system of interacting spins of a random length described by the Hamiltonian (\ref{11.1}) is sketched in 
Fig.\ref{3dIsing} {\bf (a)}. Obviously, putting all $\hat{L}_{\mathitbf r}=1$ one arrives at usual 3D Ising model shown in 
Fig.\ref{3dIsing} {\bf (b)}. In our analysis we will be interested in the case
shown in Fig.\ref{3dIsing} {\bf (c)}, when  
lattice sites are occupied by spins of two different lengths,  a part of spins 
are of length $\hat{L}_{\mathitbf r}=1$, the others being of a fixed length $\hat{L}_{\mathitbf r}=s$. When the concentration of spins of length 1 is 
$c$ and the concentration of spins of length $s$ is $1-c$, the corresponding probability distribution 
is given by:
\begin{equation}\label{11.3}
p(\hat{L}_{\mathitbf r})= c \delta (\hat{L}_{\mathitbf r}-1) +
(1-c) \delta (\hat{L}_{\mathitbf r}-s)\, .
\end{equation}
It is straightforward to see that the widely studied 
diluted Ising model, in which part of the lattice sites are occupied by Ising spins and  the other being non-magnetic
(or empty), as shown in Fig. \ref{3dIsing} {\bf (d)} is included in  the definition of $p(\hat{L}_{\mathitbf r})$ by putting $s=0$ in Eq. (\ref{11.3}).

In the case of quenched structural disorder, using the replica trick to average the disorder-dependent free energy over different configurations of spins with varying lengths one arrives at the following effective Hamiltonian \cite{Dudka23}, see also Appendix A:
\begin{eqnarray}\nonumber
{\cal H}_{\rm eff} &=& \frac{1}{2}\sum_{\mathitbf k}\sum_{\alpha=1}^n\big (\frac{1}{\beta \nu(k)}-\langle \hat{L}^2 \rangle \big ) {\phi}^\alpha_{{\mathitbf k}}
{\phi}^\alpha_{{\mathitbf{-k}}} \\ \label{11.6} &{+} &
 \sum_{\mathitbf r} \Big ( \frac{g_{1,0}}{4!}  \sum_{\alpha,\beta=1}^n  (\phi^\alpha_{\mathitbf r})^2
(\phi^\beta_{\mathitbf r})^2 {+}
\frac{g_{2,0}}{4!} \sum_{\alpha=1}^n (\phi^\alpha_{\mathitbf r})^4 \Big )   ,
\end{eqnarray}
where $\beta$ is an inverse temperature, $\nu(k)$ is the Fourier transform of the interaction potential $J(r)$,  and $n$ is a replica index. The coefficients of the different  $\phi^4$ terms can be expressed 
in terms of the moments of the random variable $\hat{L}$
resulting in:
\begin{eqnarray}\label{11.7a}
g_{1,0} &=& - 3{c(1-c)(1-s^2)^2}\, , \\ \label{11.7b}
g_{2,0} &=& 2\left({c+(1-c)s^4}\right)\, ,
\end{eqnarray}
see Appendix  \ref{eff_Ham} for details.

There are two lessons to learn at this point.
The first, and an obvious one, is that symmetry of the 
effective Hamiltonian (\ref{11.6}) is the same as the one for the diluted Ising model,
widely studied before by various approaches, see, e.g., Refs. \cite{Folk03,Pelissetto02,Holovatch02}
for reviews. This, in turn leads to the conclusion that
the asymptotic critical behavior of the random mixture of two Ising magnets and that of the Ising
magnet diluted by a non-magnetic isomorph  coincide \cite{Dudka23}. 
In particular, all renormalization group calculations performed so far to analyze 
asymptotic behavior of the diluted Ising model are directly applicable to our model too. 

The second lesson can be obtained by observing, how the coefficients $g_{0,i}$, see Eqs. (\ref{11.7a}) and
(\ref{11.7b}) depend on the concentration $c$ and the difference in spin length $s$. 
Indeed, changing the values of $c,\, s$ one can trigger the ratio
\begin{equation}\label{11.8}
 r(c,s)= \frac{g_{1,0}}{g_{2,0}}=-\frac{3}{2} \frac{c(1-c)(1-s^2)^2}{c+ (1-c)s^4}\, .
 \end{equation}
This ratio defines the
starting conditions for the renormalization group flow that governs the  approach of
the couplings $g_i$ to their fixed-point values upon renormalization. In Sec.  \ref{analyticalresults}
we will study this effect in more detail to select appropriate parameter sets ($c$ and $s$) and further
exploit them in the numerical simulations presented in Sec. \ref{MCsimulations}.

\subsection{Renormalization group flows and effective critical exponents} \label{analyticalresults}

In the following, we provide a brief explanation of the application of the field-theoretical RG approach to the effective Hamiltonian (\ref{11.6})  in the continuous limit (see Eq.~(\ref{11.9}) in Appendix~\ref{RG_desc}) in order to select the model parameters that will be used as input for the MC simulations in Sec. \ref{MCsimulations}.

Within the RG approach, a change in couplings $g_1$, $g_2$ (starting from the initial values given by Eqs. (\ref{11.7a}) and  (\ref{11.7b})) 
under the renormalization is described by RG $\beta$-functions via the flow equations \cite{Amit,ZinnJustin02,Kleinert01}:
\begin{equation} \label{11.10}
	\ell\frac{d}{{d} \ell}g_1(\ell){=}\beta_{g_1}\!\left(g_1(\ell),g_2(\ell)\right),\,\,
	\ell\frac{d}{{d}\ell}g_2(\ell){=}\beta_{g_2}\!\left(g_1(\ell),g_2(\ell)\right),
\end{equation}
where $\ell$ is the RG flow parameter. 

The flow parameter can be related by means of a so-called `matching
condition' to the correlation length, depending on the temperature distance $\tau$ to the critical point, see e.g., Ref. \cite{Folk06}. In a simpler case, it is proportional to the inverse squared correlation length, therefore the limit
$\tau\to 0$ corresponds to $\ell\to 0$. In this limit, $g_1(\ell)$ and $g_2(\ell)$ attain their
stable fixed point (FP) values  \cite{Amit,ZinnJustin02,Kleinert01}.

In the RG scheme, the effective critical exponents are calculated in the region,
where the couplings $(g_1(\ell),g_2(\ell))$  being solutions of the flow equations (\ref{11.10}) have not yet reached their FP values
and depend on $\ell$. For this purpose the RG $\gamma$-function used to calculate asymptotic critical exponents are utilized. In particular, for the correlation length and pair correlation function (anomalous dimension of the field) exponents $\nu_{\rm eff}$ and $\eta_{\rm eff}$ one
gets \cite{Folk03}:
\begin{eqnarray}\label{gam}
	\nu_{\rm eff}(\tau)&=& \frac{1}{2+{ \gamma}_{m^2}
		[g_1\{\ell(\tau)\},g_2\{\ell(\tau)\}]}+\dots,\nonumber\\ \eta_{\rm eff}(\tau)&=&{2{ \gamma}_{\phi}
		[g_1\{\ell(\tau)\},g_2\{\ell(\tau)\}]}+\dots\, ,
\end{eqnarray}
where RG functions $\gamma_{\phi}$ and ${\gamma}_{m^2}$  are described in Appendix~\ref{RG_desc}, the dots denote contribution proportional to the $\beta$-functions, 
coming from the change in the amplitude part of corresponding quantities. Usually, these parts can be neglected \cite{Folk06}.

Since our model and the site-diluted Ising model share effective Hamiltonian (see Sec. \ref{model2lengths}) we can use the RG equations of the latter.
In particular, in our calculations, we have used field-theoretical RG functions calculated in the minimal 
subtraction scheme of renormalization within the highest accessible six-loop approximation 
\cite{Kompaniets21,Adzhemyan19}. They were obtained in the form of perturbative series in 
couplings and are known to be asymptotic at best. 

Therefore, one needs to apply certain resummation 
techniques trying to  restore their convergence and to extract reliable numerical 
estimates on their basis. The results discussed below are obtained on the basis of 
conformal Borel resummation \cite{LeGuillou77} procedure generalized for a case of several couplings
\cite{Pelissetto00,Alvarez00,Pelissetto04}. It is briefly described in the Appendix \ref{RG_desc}, see also Ref. \cite{Dudka23} for more
details.

\begin{figure}[h!]
    \begin{center}
        \includegraphics[width=0.35\paperwidth]{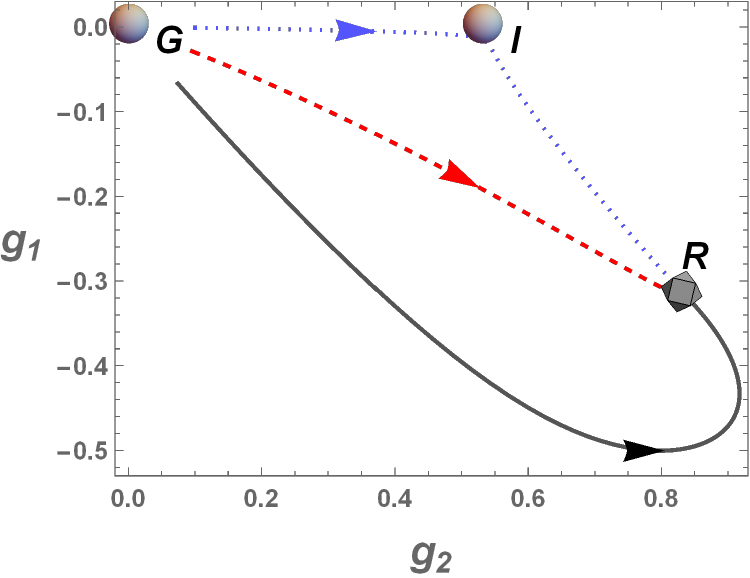}
        \caption{Renormalization group flow in the space of couplings for different initial conditions. From up to down: $r=g_1(\ell_0)/g_2(\ell_0)$= -0.01 (dotted blue), -0.3 (dashed red),  -0.9 (solid black). Unstable Gaussian ({\bf G}) and Ising ({\bf I}) FPs are shown by gray discs while a stable random Ising ({\bf R}) FP is shown by the gray diamond. \label{flow}}
    \end{center}
\end{figure}

As it is well established by now, the RG flow equation for the effective Hamiltonian (\ref{11.9}) is characterized by  
three FPs in the region $g_1\leq 0$, $g_2\geq 0$: Gaussian FP {\bf G} ($g_1=0$, $g_2=0$), pure Ising model FP {\bf I} ($g_1=0$, $g_2\neq 0$) and random Ising model FP {\bf R} ($g_1\neq 0$, $g_2\neq 0$). They are given in Fig.~\ref{flow}. Only FP {\bf R} is stable. The values of the asymptotic critical exponents we have obtained at the FP {\bf R}, used the outlined procedure, are: $\nu=0.676$ and  $\eta=0.032$. As one can see from Table \ref{table:data}, they are very close to the latest analytical estimates, see also  the review in Ref. \cite{Kompaniets21} for earlier data.

Now let us proceed analyzing the RG picture in the non-asymptotic regime, to see how the 
effective critical exponents approach their asymptotic values depending on the initial 
couplings of the effective Hamiltonian. 

Since we aim to verify analytic predictions by 
the MC simulation in the following section, let us present several effective exponents 
for selected initial conditions, which will serve as input for MC calculations afterwards.

To do that, we solve numerically the system of differential equations (\ref{11.10}) with resummed 
$\beta$-functions getting  the running values of the couplings.
As already mentioned, the properties of the RG flow depend on the initial conditions ($g_1(\ell_0)$,
$g_2(\ell_0)$) of the differential equations (\ref{11.10}). We have chosen ($g_1(\ell_0)$,
$g_2(\ell_0)$) in the vicinity of FP {\bf G} for three distinct values $r=g_1(\ell_0)/g_2(\ell_0)$. 

As one can see from
the Fig.~\ref{flow}, all three flows lead to the stable FP with the decrease of $\ell$. However, their behavior 
strongly depends on the initial conditions. For  $r=-0.01$,  the flow first stays within the basin 
of attraction of the unstable Ising FP (blue dotted curve in Fig.~\ref{flow}). For $r=-0.3$, the flow 
directly approaches the stable random FP (red dashed curve), while for $r=-0.9$, the running values of the couplings approach their stable FP values with 'overshooting' behavior (black curve).

\begin{figure} [h!]
    \begin{center}
        \includegraphics[width=0.35\paperwidth]{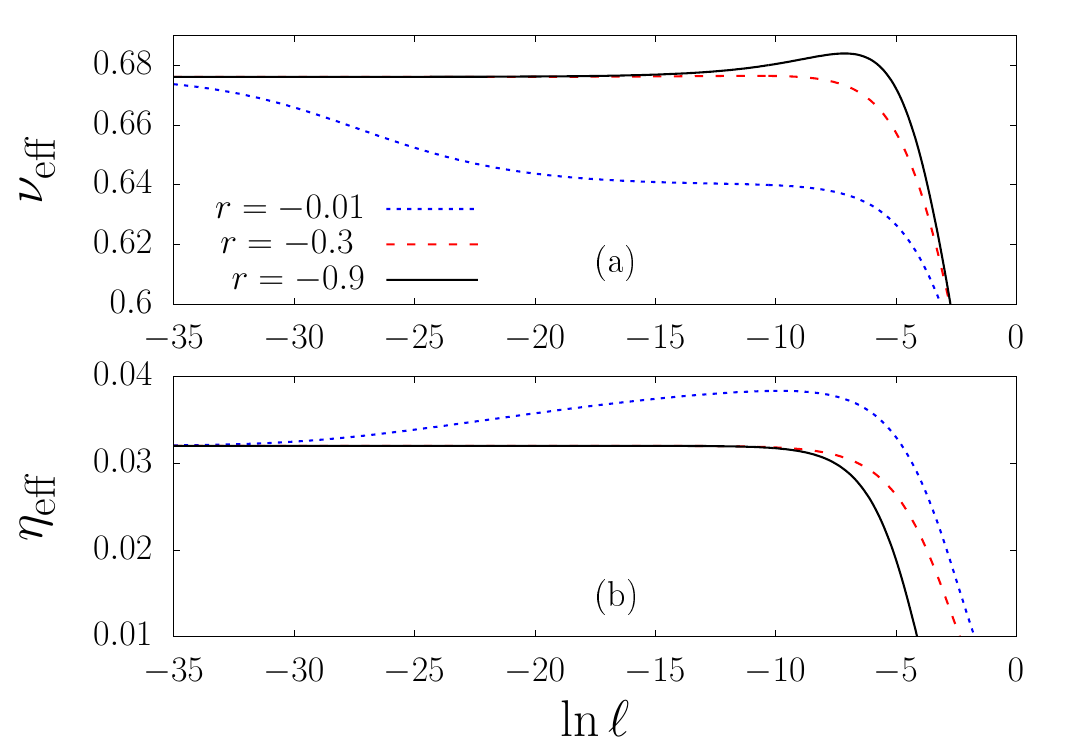}
        \caption{Dependencies of  the
        correlation length effective critical exponent $\nu_{\rm eff}$ (top, panel (a)) and the
        pair correlation function effective critical exponent $\eta_{\rm eff}$  (bottom, panel (b)) on the flow parameter
        calculated along the RG flows of Fig. \ref{flow}. Notice that we have used the same colors as in Fig. \ref{flow}. \label{exponn}}
    \end{center}
\end{figure}

The dependence of effective critical exponents $\nu_{\rm eff}$ and $\eta_{\rm eff}$  on the flow parameter $\ell$  is presented in Fig.~\ref{exponn}. 
Depending on the initial conditions, the
effective critical exponent $\nu_{\rm eff}$ either reaches its universal values at the
stable FP comparatively fast (red curve), or attains the values that
differ from the stable FP ones in a broad crossover region governed  by the pure Ising model
universality class critical exponents (blue curve), or its value exceeds the stable FP ones 
(peak in the black curve). We have not observed the third case for $\eta_{\rm eff}$. 
A possible explanation is that the value of $\eta$ is small, therefore  the approximation 
neglecting contributions marked by dots in Eq. (\ref{gam}) is not relevant to its calculation.
 
In the following section, we will check our theoretical predictions for the effective critical exponents for the flows 
with initial conditions $r=-0.3$ and $r=-0.9$. One can see from Eq.(\ref{11.8}) that one option for the first case can be chosen as $s=1.7$, $c\approx 0.53$, while for the second case, one can choose  $s=3$, $c\approx 0.79594$.

 \section{Monte Carlo simulations \label{MCsimulations}}

In this section, we will describe the numerical methodology and the observables we have used in our analysis. After that, we will show our characterization of the critical properties of the model for two sets of model parameters. 

\subsection{The Model} \label{MCModel}

Since we will consider the nearest neighbor interaction, it is convenient 
to re-write the Hamiltonian
(\ref{11.1}) as
\begin{equation}
\label{eq:Hamiltonian}
{\cal H}=-\sum_{<\mathitbf{r},\mathitbf{r}^\prime>} J_{\mathitbf{r},\mathitbf{r}^\prime} S_\mathitbf{r} S_{\mathitbf{r}^\prime}
\end{equation}
where the sum runs over all the nearest neighbors pairs in a three-dimensional cubic lattice of size $L$ (and volume $V=L^3$) with  periodic boundary conditions.  The variables $S_{\mathitbf{r}}$ are Ising ones ($\pm 1$).
$J_{\mathitbf{r},\mathitbf{r}^\prime}$ are quenched variables (disorder) given by
\begin{equation} \label{interact}
J_{\mathitbf{r},\mathitbf{r}^\prime}=\hat{L}_{\mathitbf{r}} \hat{L}_{\mathitbf {r}^\prime}
\end{equation}
We recall that due to the distribution function (\ref{11.3}) $\hat{L}_\mathitbf{r}$ is 1 with probability $c$, otherwise $s$. 
Notice since all $J_{\mathitbf{r},\mathitbf{r}^\prime}>0$ we can use cluster methods to simulate this model. 

\subsection{Observables} \label{Observables}

Let us introduce the magnetization per spin
\begin{equation}
{\mathcal M} =\frac{1}{V}\sum_\mathitbf{r}
S_\mathitbf{r} \, .
\label{eq:mag}
\end{equation}

Then the susceptibility can be written as
\begin{equation}
\chi=V\overline{\left\langle {\mathcal M}^2 \right\rangle}\, ,
\label{susceptibility}
\end{equation}
and the Binder cumulant as
\begin{equation}
U_4=1-\frac{1}{3}\frac{\overline{\langle {\mathcal M}^4\rangle}}
           {\overline{\langle{\mathcal M}^2 \rangle}^2} \, ,
\label{binder}
\end{equation}
where $\langle(\cdot \cdot \cdot)\rangle$ is the thermal
average, $\overline{(\cdot \cdot \cdot)}$ is the average over
the disorder (length of the spins). 
In the thermodynamic limit $V\to\infty$, the Binder cumulant will be zero in the paramagnetic phase 
and will take the value 2/3 in the ferromagnetic one.

A very useful way to characterize the phase transition is the use of the  correlation length on a finite lattice~\cite{XIL,Amit} defined as
\begin{equation}
\xi=\left(\frac{\chi/F-1}{4\sin^2(\pi/L)}\right)^\frac{1}{2}\,,
\label{XI}
\end{equation}
with
\begin{equation}
F=\frac{V}{3}\overline{\left\langle |\mathcal{F}(2\pi/L,0,0)|^2+\mathrm{two~permutations}\right\rangle}\, ,
\end{equation}
${\cal F}$ being  the Fourier transform of the
magnetization 
\begin{equation}
  {\cal F}(\mathitbf{k})=
  \frac{1}{V}\sum_{\mathitbf{r}} e^{\mathrm i
\mathitbf{k}\cdot\mathitbf{r}} S_\mathitbf{r} \, .
\end{equation}
Defined by Eq. (\ref{XI}), $\xi$ is only a true correlation length in the paramagnetic phase, therefore, $R_\xi\equiv \xi/L$ goes to zero in this phase as $L$ diverges. $\xi$ will diverge in the ferromagnetic phase as $\xi\sim L^{d/2}$ and consequently $R_\xi$ (in three dimensions). Therefore, the 
curves of $R_\xi$ for different lattice sizes will cross near the critical temperature~\cite{XIL,Amit}.

To obtain the correlation length $\nu$ exponent from the analysis of $R_\xi$ we need to estimate the derivative
of the correlation length with respect to the inverse temperature $\beta=1/T$, $\partial_\beta \xi$, by 
computing connected average values of different moments of
the magnetization (and of $\cal F$ in the case of $\xi$).  The derivative
for quantity $\overline{\langle {\cal O}\rangle}$ is calculated using the following relation
\begin{equation}
\partial_\beta \overline{\langle {\cal O}\rangle}=
 \overline{\partial_\beta\langle {\cal O}\rangle}=
\overline{\left\langle_{\vphantom{|}} {\cal{O H}} - \langle{\cal O}
 \rangle \langle {\cal H} \rangle\right\rangle}.
\end{equation}

To correct the bias induced by a small number of measurements in each sample
we have applied a third-order extrapolation as described in Ref.~\cite{Extra}.

The $g_2$ cumulant,  measuring the lack of self-averageness of the system, is given by
\begin{equation}
g_2=\frac{ \overline{\langle{\mathcal M}^2  \rangle^2}-\overline{\langle
    {\mathcal M}^2  \rangle}^2}{\overline{\langle {\mathcal M}^2  \rangle}^2}\,.
\end{equation}

The susceptibility will be a self-averaging quantity if $g_2$ tends to zero as $L$ increases. 
If it is not the case, the susceptibility does not self-average (for example, see
Refs.~\cite{Wiseman95,Aharony96}).

\subsection{Simulation parameters} \label{MCparameters}

The model has been studied by means of numerical simulations. We have used a combination of the Wolff single-cluster algorithm~\cite{Wolff:89} complemented by Monte Carlo local updates. In particular, our elementary Monte Carlo step (one sweep) consisted of $L$ single-cluster updates, followed by a sequential full-lattice Metropolis update~\cite{Ballesteros98,Extra}. 

In addition, we have assessed the thermalization for each temperature monitoring against the logarithm of the Monte Carlo time non-local observables such as the susceptibility or the cumulant $R_\xi$. In Table \ref{tab:parametersA} we report the parameters of our numerical simulations.

\begin{table}[ht!]
\caption{Parameters used in the numerical simulations. $N_{\rm samples}$ is the number of
  disorder realizations and $N_{\rm sweeps}$ is the number of the elementary Monte Carlo steps (built by $L$ single-cluster updates, followed
by a sequential full-lattice Metropolis update). Finally, the length of the spins is 1 with probability $c$ and $s$ with probability $1-c$. } 

\centering
\label{tab:parametersA}
\begin{tabular}{|c||c|c|c|c||c|c|c|c|}
\hline
$L$  &$s$ & $c$ & $N_\text{samples}$ & $N_\text{sweeps}$ & $s$ & $c$ & $N_\text{samples}$ & $N_\text{sweeps}$ \\
\hline
8 & 1.7  & 0.53 & 26000 & 256 & 3.0 & 0.795943 & 48440 & 256 \\
12 & 1.7 & 0.53 & 26000 & 256 & 3.0 & 0.795943 & 48950  & 256\\
16 & 1.7 & 0.53 & 26000 & 256 & 3.0 &  0.795943 & 49000  & 256 \\
24 & 1.7 & 0.53 & 36000 & 256 & 3.0 & 0.795943 & 49000  & 256 \\
32 & 1.7 & 0.53 & 26070 & 256 & 3.0 & 0.795943 & 48996  & 256\\
48 & 1.7 & 0.53 & 28080 & 256 & 3.0 & 0.795943 & 48997  & 256\\
64 & 1.7 & 0.53 &  31440 & 256 & 3.0 & 0.795943 & 41725 & 256\\
\hline
\hline
\end{tabular}
\end{table}

In particular, we have simulated two sets of parameters: $(s=1.7,c=0.53)$  for which the theory ($r=-0.3$) predicts small scaling corrections (almost a perfect action) and  $(s=3.0,c=0.795943)$  for which the prediction of the theory ($r=-0.9$) is the appearance of huge scaling corrections. Below,  we will refer to the two simulated cases just as $s=1.7$ and $3$. Notice that we have computed the corresponding RG flows and effective exponents for these two values of $r$, see Fig. \ref{exponn}.

To simulate different temperatures we have performed an annealing from the highest temperature to the lowest one. For $s=3$ we have simulated in the annealing procedure 20 temperatures in the critical region, for $s=1.7$ we have simulated 20 temperatures for the smaller lattice sizes ($L\le 32$) and 7 for $L=48$ and 64. 
We have used a fifth-order polynomial-based analysis to compute the crossing temperatures.

Finally, we have estimated  the statistical errors using the jackknife method~\cite{young:15}.

\subsection{Results} \label{MCresults}

\begin{figure}[h]
    \begin{center}
        \includegraphics[width=0.35\paperwidth]{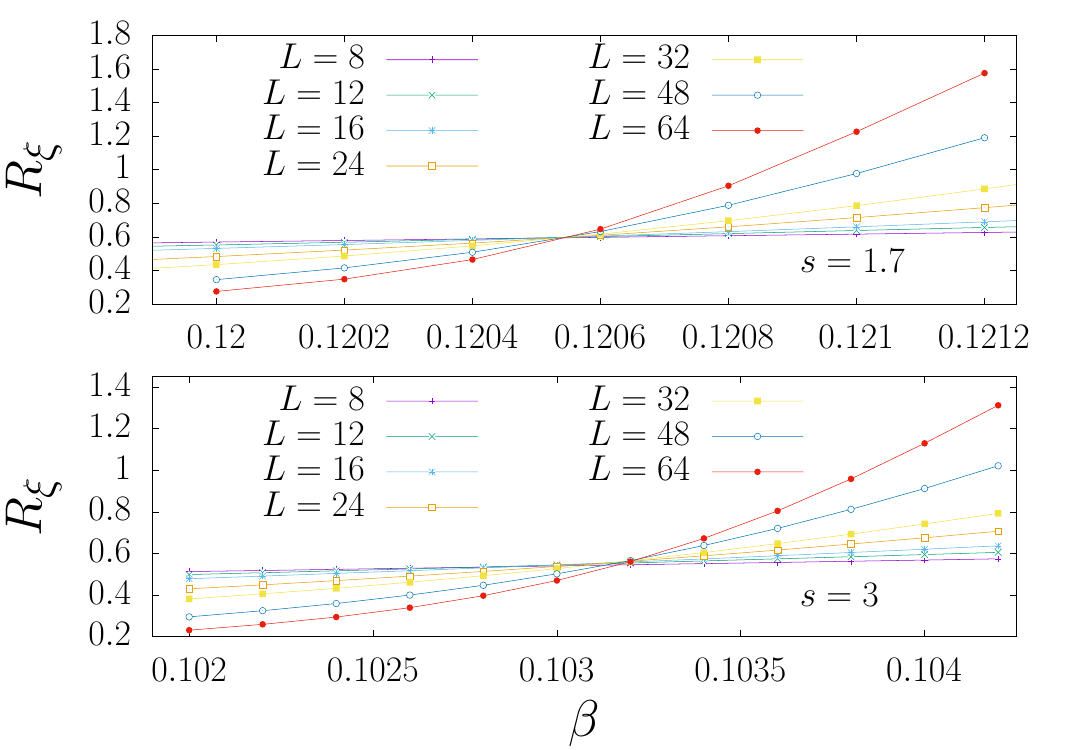}
        \caption{Behavior of $R_\xi$ versus the inverse temperature $\beta$ for $s=1.7$ (top) and $s=3$ (below). The crossing point of the different curves marks the effective transition point. \label{fig:crossing}}
    \end{center}
\end{figure}

We have focused on the cumulant $R_\xi$ to characterize the phase transition for both sets of parameters $s$ and $c$. In Fig. \ref{fig:crossing} we show the curves $R_\xi(\beta)$ ($\beta=1/T$)  for the different lattice sizes and two values of $s$. The phase transitions are very clear for both values of $s$ (the different curves cross at the apparent critical inverse temperature).

We have analyzed these two phase transitions (for $s=1.7$ and 3) using the quotients method
\cite{nightingale:76,ballesteros:96,Amit}, which is based on the analysis of the different observables on the crossing point of the curves $L$ and $2 L$, in this case of the cumulant $R_\xi$ (we describe briefly this method in Appendix \ref{quotient}). The intermediate data of this analysis based on the quotients method is reported in Table \ref{tab:resultsQA} (based on the crossing points of $R_\xi$) and \ref{tab:resultsCumulantsQA} (based on the crossing points of $R_{g_2}$ and $R_{U_4}$).

\begin{table*}[hbt!]
\caption{Quotients method results for the two simulated values of $s=3$ and $1.7$ from the crossing points of $R_\xi$ for lattice sizes $L_1$
  and $L_2$.}  \centering
\label{tab:resultsQA}
\begin{tabular}{|c||c|c|c|c|c|c|c|}
\hline
$s$ & $L_1/L_2$  & $\beta_\text{cross}$ & $R_\xi$ & $\nu_\xi$  &
  $\eta$ & $Q_{U_4}$\\
\hline\hline
1.7 & 8/16  & 0.12051(1) & 0.594(1) & 0.694(2) & 0.030(4)  & 0.984(1)  \\
1.7 & 12/24 & 0.120552(6)& 0.598(1) & 0.684(2) & 0.031(4)  & 0.991(1)  \\
1.7 & 16/32 & 0.120550(4)& 0.599(1) & 0.680(2) & 0.033(4)  & 0.992(1)  \\
1.7 & 24/48 & 0.120549(2)& 0.598(1) & 0.679(2) & 0.033(4)  & 0.995(1)  \\
1.7 & 32/64 & 0.120555(1)& 0.601(1) & 0.678(2) & 0.032(4)  & 0.995(1)  
\\\hline\hline
3.0 & 8/16  & 0.10289(1) & 0.537(1)  & 0.757(3) & 0.059(6) & 0.982(2)   \\ 
3.0 & 12/24 & 0.1031017) & 0.550(1)  & 0.735(2) & 0.044(5) & 0.988(2)   \\ 
3.0 & 16/32 & 0.103156(7) & 0.556(1) & 0.725(2) & 0.039(5) & 0.989(1)    \\
3.0 & 24/48 & 0.103185(4) & 0.561(1) & 0.717(2) & 0.036(5) & 0.988(2)   \\
3.0 & 32/64 & 0.103212(2) & 0.567(1) & 0.713(1) & 0.034(6) & 0.989(2)   
\\\hline\hline
\end{tabular}
\end{table*}

\begin{figure}[h!]
    \begin{center}
        \includegraphics[width=0.35\paperwidth]{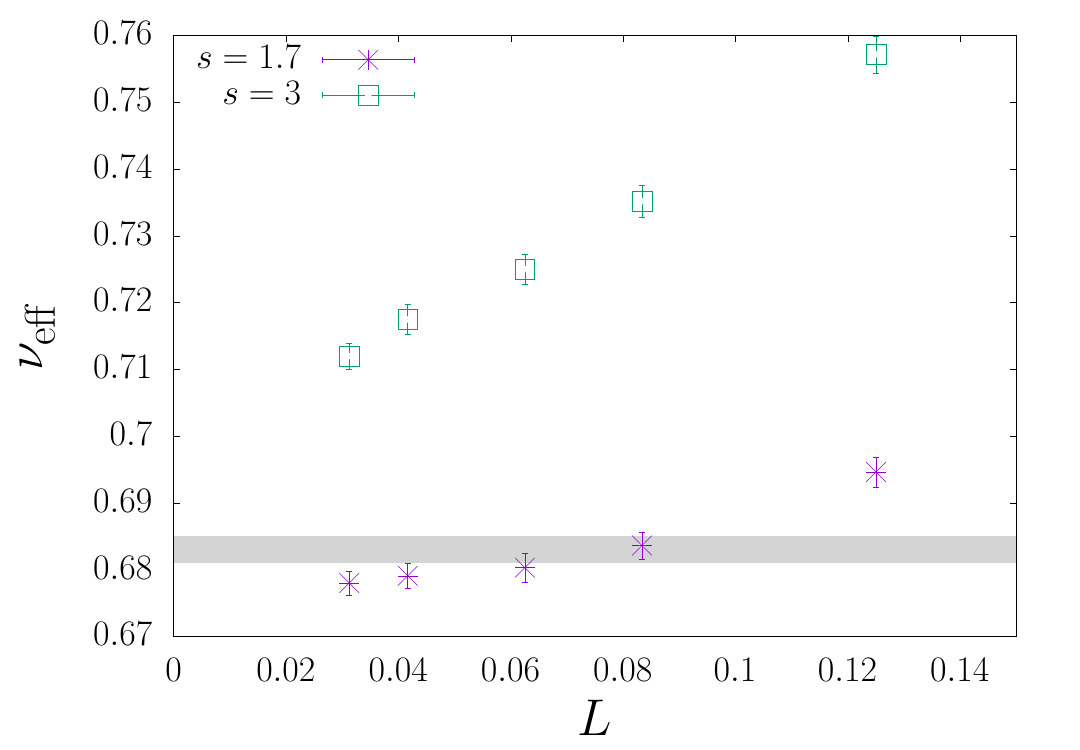}
        \caption{ Behavior of the effective exponent $\nu(L)$ computed with the quotients method as a function of $1/L$ for the two simulated cases $s=1.7$ and $3.0$. The shadow horizontal region is the most accurate value from numerical simulations~\cite{Hasenbusch07} (one standard deviation).\label{fig:nueff}}
    \end{center}
\end{figure}

\begin{figure}[h!]
    \begin{center}
        \includegraphics[width=0.35\paperwidth]{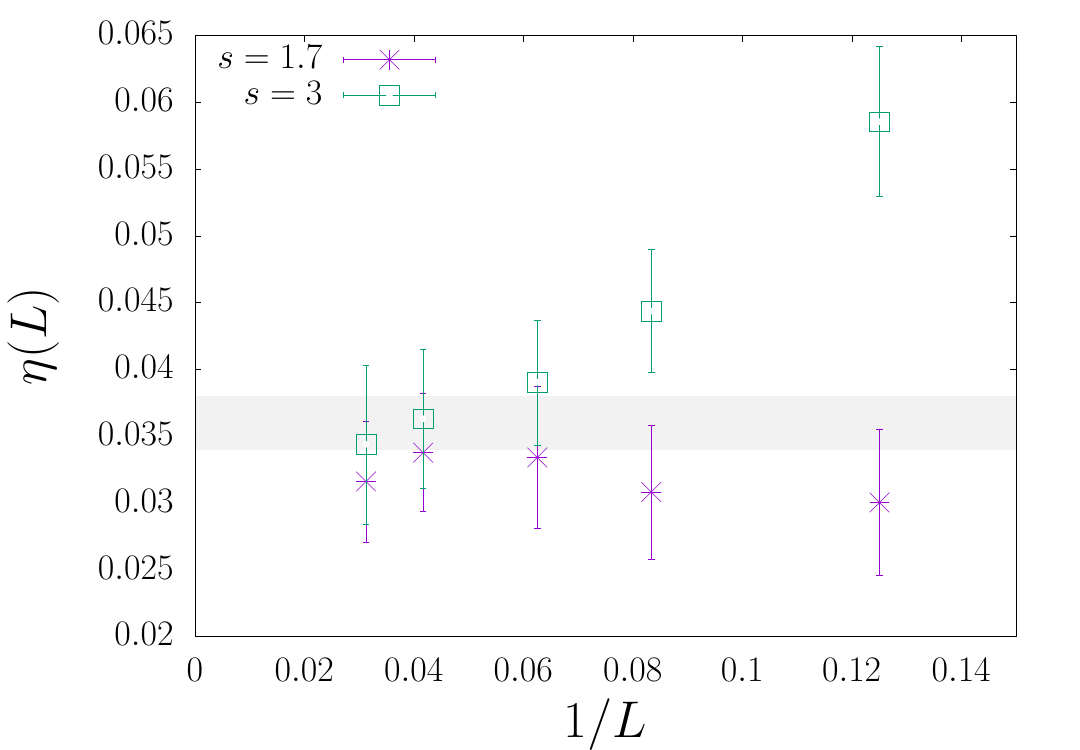}
        \caption{Behavior of the effective exponent $\eta(L)$ computed with the quotients method as a function of $1/L$ for the two simulated cases $s=1.7$ and $3.0$. The shadow horizontal region is the most accurate value from numerical simulations~\cite{Hasenbusch07} (one standard deviation). \label{fig:etaeff}}
    \end{center}
\end{figure}

One can take a much more detailed look at the data for the effective $\nu$ and $\eta$ exponents by plotting them as a function of $1/L$ in order to check their asymptotic behavior, see Figs. \ref{fig:nueff} and \ref{fig:etaeff}, where we have also added the most accurate values from the 3D site-diluted Ising model~\cite{Hasenbusch07}. Notice that the $s=1.7$-case is almost a perfect action (as predicted in Sec. \ref{analyticalresults} by our RG analysis) and for small values of $L$ falls on the asymptotic value for both exponents. On the contrary, and again as predicted by our RG calculations, the $s=3$-case shows stronger scaling corrections.

A more quantitative analysis of the asymptotic values of the exponents $\nu$ and $\eta$, the correction-to-scaling exponent and the asymptotic values of the crossing points of $R_\xi$ and $U_4$ (which are also universal) are presented in Table \ref{table:extraQA}. To obtain these asymptotic values, we have analyzed the effective  $\nu$ and $\eta$ exponents using Eq. (\ref{eq:QO}) and the values of $R_\xi$, $g_2$ and $U_4$ using Eq. (\ref{eq:QOB}) (see Appendix \ref{quotient}). In all the cases presented in this paper, the statistical 
qualities of the different fits are very good (the $p$-values of the $\chi^2$-fits are bigger than 
5\% in all the fits).

In particular, in Table \ref{table:extraQA}, we have extrapolated  $\beta_c$, $\nu$, $\eta$, $g_2$, $R_\xi$ and $U_4$ using a three-parameter fit, so, the $\omega$ exponent was not fixed finding that $\omega$ 
presented a large error.  To compute the $\omega$ exponent we have analyzed the scaling of $Q_{U_4}$ (see Eq. (\ref{eq:dimensionless-quotients})).

Notice that our final results for the universal quantities (see Table \ref{table:extraQA}) present a very good agreement (differences of less than 1.8  standard deviations in the worst case) with all the universal quantities of the 3D site-diluted Ising model, see Table \ref{table:data}. 

It is important to notice that in the $s=1.7$-case we are obtaining a correction-to-scaling exponent that is very compatible with the sub-leading one of the 3D diluted Ising model: our computed value is $\omega=0.94(15)$ versus $\omega_2=0.82(8)$, and the difference between these two exponents is within one standard deviation. However, in the $s=3.0$-case the computed $\omega$-exponent is compatible with the leading one: our computed value is $\omega=0.31(12)$ versus $\omega_1=0.33(3)$, again, the difference between these two exponents is within one standard deviation.

\begin{figure}[h!]
    \begin{center}
        \includegraphics[width=0.35\paperwidth]{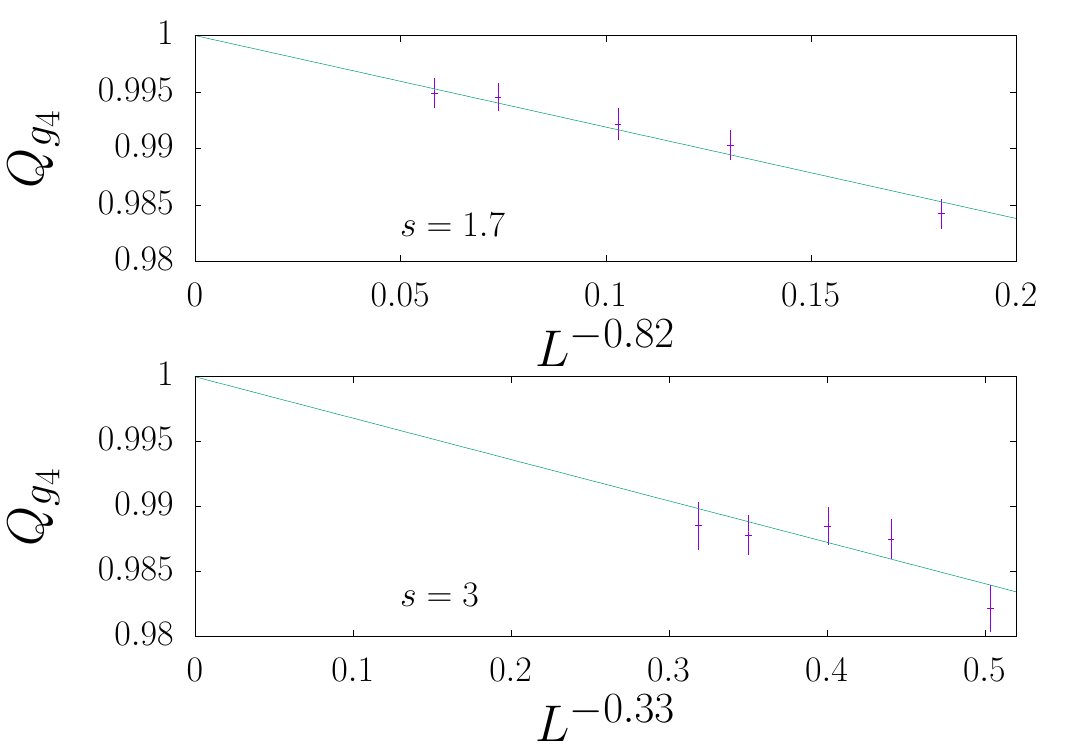}
        \caption{Behavior of $Q_{g_4}$ computed with the quotients method as a function of $1/L^{0.82}$ (top, $s=1.7$-case) and $L^{-0.33}$ (bottom, $s=3.0$-case). This observable, in a second order phase transition, always extrapolates to 1 (see Appendix \ref{quotient}). Notice that we have assumed the leading and sub-leading correction-to-scaling exponents of the 3D diluted Ising model (see Ref.~\cite{Hasenbusch07}). The straight lines are linear fits to the function $f(x)=1+a_s x$ where $a_s$ depends on $s$: the statistical quality of both fits is very good (see the main text).   \label{fig:Q4}}
    \end{center}
\end{figure}

To emphasize this point, we report the statistical quality of the fits of $Q_{g_4}$ against $1+ a_s/L^{\omega_s}$, where $\omega_{s=3.0}=0.33$ and $\omega_{s=1.7}=0.82$ (the leading and subleading correction-to-scaling exponents as computed in Ref. \cite{Hasenbusch07} for the 3D diluted Ising model) are very good (for the two simulated values of $s$). In particular, we can quote the values of the goodness of fit $\chi^2/\mathrm{d.o.f}=2.4/4$ ($p$-value=$83.5\%$) for $s=1.7$ and
$\chi^2/\mathrm{d.o.f}=3.85/4$ ($p$-value=$44.8\%$)) (see also Fig. \ref{fig:Q4}). Note that d.o.f denotes the number of degrees of freedom of the fit \cite{young:15}.\footnote{Furthermore, if we try $\omega=0.33$ in the $s=1.7$-case, we obtain a very bad fit: $\chi^2/\mathrm{d.o.f}=18.3/4$ ($p$-value=$0.1\%$). We also obtain a very bad fit for $s=3.0$ assuming $\omega=0.82$: $\chi^2/\mathrm{d.o.f}=12.8/4$ ($p$-value=$0.2\%$). These results do not change even taking into account the error bars on $\omega$, for example by doing fits using $\omega=0.36$ and $\omega=0.74$.}

\begin{table*}[hbt!]
\caption{Quotients method results ($s=3$ and $1.7$) for the values
of the cumulants $g_2$ and $U_4$ at the crossing points of $R_{g_2}$ and  $R_{U_4}$ (respectively) for lattice sizes $L_1$
  and $L_2$.}  \centering
\label{tab:resultsCumulantsQA}
\begin{tabular}{|c||c|c|c|}
\hline
$s$ & $L_1/L_2$ & $U_4$ & $g_2$ \\
\hline\hline
1.7 & 8/16   & 0.4660(9)   & 0.113(1)\\
1.7 & 12/24  &  0.4616(9)  & 0.122(1) \\
1.7 & 16/32  &  0.4591(9)  & 0.127(1)\\
1.7 & 24/48  & 0.4559(7)   & 0.130(1)\\
1.7 & 32/64  &  0.4563(9)  & 0.134(2)\\
\hline\hline
3.0 & 8/16  & 0.4525(9) & 0.238(2)\\
3.0 & 12/24 & 0.456(1)  & 0.242(3) \\
3.0 & 16/32 & 0.457(1)  & 0.237(3) \\
3.0 & 24/48 & 0.457(1)  & 0.222(3)\\
3.0 & 32/64 & 0.459(1)  & 0.215(2) \\
\hline\hline
\end{tabular}
\end{table*}

\begin{table*}[hbt!]
\caption{Extrapolated results, using the quotients method,  for $s=1.7$ and $s=3.0$.}  \centering
\label{table:extraQA}
\begin{tabular}{|c||c|c|c|c|c|c|c|}
\hline 
$s$ & $\nu$  & $\eta$ & $\omega$ &  $R_\xi$ & $U_4$ & $g_2$ & $\beta_c$ \\
\hline \hline
1.7 & 0.678(2) & 0.033(7) & 0.94(15) &  0.5990(8) & 0.453(3)  & 0.138(5) & 0.12056(1) \\
\hline 
3.0 & 0.706(6) & 0.033(7) & 0.31(12)  & 0.579(8) & 0.46(3)  &   0.13(1) & 0.1033(1)\\
\hline \hline
\end{tabular}
\end{table*}

\section{Conclusions} \label{discussion}

In this paper, we have presented analytical and numerical analysis of the 3D Ising model in the presence of a disorder-inducing spin length variable providing a new view on how structural disorder affects magnetic phase transitions. 

We have computed the critical exponents and cumulants of this model for two values of the parameters, finding very good agreement with the critical exponents and cumulants of the 3D site-diluted Ising model as predicted by a perturbative field-theoretical RG analysis. This is the main result of the paper.

Furthermore, we have found that the  two simulated models, with respect to their scaling corrections, behave as suggested by the perturbative field-theoretical RG analysis. We considered the limiting case for the model when spins can take two different lengths $\hat{L}=1$ and $\hat{L}=s$. Based on a six-loops RG analysis for effective exponents, we simulated the critical behavior for the system with two sets of parameters: $s=1.7$ (with concentration $c=0.53$) and $s=3$ (with concentration $c=0.79594$). 
The perturbative RG analysis provides different behaviors for the scaling corrections. Using the quotients method, we observed distinct and strong differences in the scaling corrections for each case. Namely, for spin length $s=1.7$ the obtained values align closely with the predicted asymptotic values for the critical exponents, indicating minimal scaling corrections, whereas for $s=3$ case the strong scaling corrections appear.

However, the quantitative behavior provided by the six-loop calculation as a function of the renormalization scale does not correctly describe the behavior of the effective critical exponents computed in the numerical simulations (even for some cases, it misses the right sign of the scaling corrections). It is clear that higher orders in the perturbative expansions with, maybe, improved summations techniques should be taken into account to obtain the right quantitative behavior of the effective critical exponents.

Therefore, we have confirmed our previous theoretical results obtained from the field-theoretical RG predictions \cite{Dudka23} regarding the universality class of the model with variable spin lengths by performing extensive MC simulations. This leads to the strong conclusion that we can create new types of critical behavior in disordered Ising-like systems by considering systems with multiple magnetic components, where spin length and concentration can be used to fine-tune critical properties.

\subsection*{ACKNOWLEDGMENTS}
M.D., M.K. and Yu.H were supported by the National Research Foundation
of Ukraine Project 2023.03/0099 ``Criticality of complex systems: fundamental aspects and applications''. 
J.J.R.L was partially supported by Ministerio de Ciencia, Innovaci\'on y Universidades (Spain), Agencia Estatal de Investigaci\'on (AEI, Spain, 10.13039/501100011033), and European Regional Development Fund (ERDF, A way of making Europe) through Grant No.\ PID2020-112936GB-I00, and by the Junta de Extremadura (Spain) and Fondo Europeo de Desarrollo Regional (FEDER, EU) through Grants No.\ GR21014 and No.\ IB20079. 
We have run our simulations in the computing facilities of the Instituto de Computaci\'{o}n Cient\'{\i}fica Avanzada de Extremadura (ICCAEx).

\appendix

\section{Effective Hamiltonian of the random-spin-length Ising model} \label{eff_Ham}
When the spin of lengths $\hat{L}_{\mathitbf r}=1$ and $\hat{L}_{\mathitbf r}=s$ are randomly distributed
and fixed in a certain configuration $\{\hat{L}\}$,  physical observables are derived from the configurational average of the free energy \cite{Brout59}: 
\begin{equation}\label{11.4}
 F=-\beta^{-1} \langle \ln Z (\{\hat{L}\}) \rangle_{\{\hat{L}\}}\, , 
\end{equation}
where  $Z (\{\hat{L}\})$ is configuration-dependent partition function and the averaging over
spin-length configurations is performed with the distribution function (\ref{11.3}):
\begin{equation}
\label{11.5}
\langle \dots \rangle_{\{\hat{L}\}} = \prod_{\mathitbf r} \sum_{\hat{L}_{\mathitbf r}}p(\hat{L}_{\mathitbf r})(\dots)\, .
\end{equation}
The product in (\ref{11.5}) is performed over all lattice sites and the sum over  $\hat{L}_{\mathbf r}$ spans 
on the two values of $\hat{L}_{\mathitbf r}=1,\, s$. 

In classical settings, one resorts on the replica approach \cite{Dotsenko01} to perform the configurational 
averaging in (\ref{11.4}). Representing the logarithm of the partition function in the form:   
\begin{equation} \label{repl}
 \ln Z (\{\hat{L}\}) = \lim_{n\to 0} \frac{\big (Z (\{\hat{L}\})\big )^n -1}{n}\, , 
\end{equation}
and using Hamiltonian (\ref{11.1}) one arrives at the 
effective   Hamiltonian \cite{Dudka23} with two $\phi^4$ terms of different symmetry, as
written in Eq. (\ref{11.6}).
The coefficients $g_{1,0}$, $g_{2,0}$ at the different  $\phi^4$ terms in the effective Hamiltonian 
(\ref{11.6}) can be expressed in terms of the moments of the random variable $\hat{L}$:
\begin{eqnarray} \nonumber
g_{1,0} &=& - 3u_2^2\left(\langle \hat{L}^4 \rangle - \langle \hat{L}^2 \rangle^2 \right) \,,\\ \nonumber
g_{2,0} &=& u_4 \langle \hat{L}^4 \rangle\,,
\end{eqnarray}
with $u_2/2=1/2,\, u_4/4!=1/12$. The moments of the random variable $\hat{L}$ with  the distribution 
 function (\ref{11.3})  readily follow:
\begin{eqnarray}
\langle \hat{L}^k \rangle &=& c + (1-c)s^k\, ,\\
\langle \hat{L}^4 \rangle - \langle \hat{L}^2 \rangle^2 &=& c(1-c)(1-s^2)^2. 
\end{eqnarray}
 With these expressions at hand, we get $g_{i,0}$ dependence on $c$ 
and $s$ via Eqs. (\ref{11.7a}), (\ref{11.7b}). 

\section{Field-theoretical RG approach, RG functions and their resummation} \label{RG_desc}

In the continuous limit the effective Hamiltonian (\ref{11.6}) can be rewritten using the  notations of Ref. \cite{Kompaniets21} as follows:
\begin{eqnarray}
	{\cal H}_{\rm eff} &{=}& \!\int\!\! d^d {\mathitbf r}
	\Bigg\{\!\frac{1}{2}\sum_{\alpha{=}1}^n\left[m_0^2(\phi^\alpha_{\mathitbf r})^2{+}
	|\mathbf{\nabla}\phi^\alpha_{\mathitbf r}|^2\right]\nonumber\\ \label{11.9}
	&&+ \frac{g_{1,0}}{4!}\sum_{\alpha,\beta=1}^n  (\phi^\alpha_{\mathitbf r})^2  (\phi^\beta_{\mathitbf r})^2+
	\frac{g_{2,0}}{4!} \sum_{\alpha=1}^n  (\phi^\alpha_{\mathitbf r})^4
	\!\Bigg\},
\end{eqnarray}
where  squared bare (unrenormalized)  mass $m_0^2$
measures the distance in temperature $\tau$ to the critical point, and the ratio of
bare couplings $g_{1,0}/g_{2,0}$ equals to  $r(c,s)$, see Eq. (\ref{11.8}).

The FP $(g_1^*,g_2^*)$ of the RG transformation 
is defined by means the zero of both $\beta$-functions: $\beta_{g_1}\left(g_1^*,g_2^*\right)=0,\,\,
\beta_{g_2}\left(g_1^*,g_2^*\right)=0$. 
The FP is stable if both eigenvalues of the stability matrix 
\begin{equation}\label{11.11}
	B_{ij}\equiv  \frac{\partial \beta_{g_i}}{\partial g_j}, \hspace{3em}
	i,j=1,2,
\end{equation}
presents, in this FP,  positive real parts.

 If the stable FP
is reachable from the initial conditions -- recall that for the
effective Hamiltonian (\ref{11.6}) the initial conditions depend on
$c$ and $s$ and lie in the region $g_1\leq 0, \, g_2> 0$, -- it corresponds to
the critical point of the system. 

Furthermore, asymptotic values of the critical exponents
are defined by the FP values of the RG $\gamma$-functions. In particular, the
 correlation length and the pair correlation function (anomalous dimension of the field) exponents 
 $\nu$ and $\eta$, respectively,  are expressed in terms of the
RG functions $\gamma_{\phi}$ and ${\gamma}_{m^2}$ describing renormalization of the field $\phi$
and of the  squared mass $m^2$
correspondingly
\cite{Amit,ZinnJustin02,Kleinert01}:
\begin{equation} \label{11.12}
\nu^{-1}={2+{ \gamma}_{m^2}\left(g_1^*,g_2^*\right)},\qquad 	\eta={2}{{ \gamma}_{\phi}\left(g_1^*,g_2^*\right)}.
\end{equation}

In our calculations we have used the most up-to-date  RG functions obtained for the effective Hamiltonian (\ref{11.9}) 
with a record six-loop accuracy in the minimal subtraction renormalization scheme \cite{Kompaniets21} (see also Ref.\cite{Adzhemyan19}). The perturbative expansions for the RG functions are of the form:
\begin{eqnarray}\label{b1}
	\beta_{g_1}(g_1,g_2)&=&-{g_1} \left(\varepsilon-\frac{8 {g_1}}{3}-2 {g_2}+\frac{14 g_1^2}{3}+\frac{22 {g_1} {g_2}}{3}\right.\nonumber\\&&\left.+\frac{5 g_2^2}{3} +\dots\right),
	\\
	\beta_{g_2}(g_1,g_2)&=&-{g_2} \left(\varepsilon-4 {g_1}-3 {g_2} +\frac{82 g_1^2}{9}+\frac{46 {g_1} {g_2}}{3}\right.\nonumber\\&&\left.+\frac{17 g_2^2}{3}+\dots\right),\label{b2}
	\\
	\gamma_{\phi}(g_1,g_2)&=&\frac{g_1^2}{18}+\frac{{g_1} {g_2}}{6}+\frac{g_2^2}{12}+\dots\label{gam1}
	\\  \label{gam2}
	\gamma_{m^2}(g_1,g_2)&=&- g_2-\frac{2 g_1}{3}+\frac{5 g_1^2}{9}+\frac{5 {g_1} {g_2}}{3}\nonumber\\&&
    +\frac{5 g_2^2}{6} +\dots\,,
\end{eqnarray}
with $\varepsilon=4-d$ and the dots indicate higher order terms currently known up to $O(g_i^7)$ and $O(g_i^6)$ for
the $\beta$- and $\gamma$-functions correspondingly \cite{Kompaniets21,Adzhemyan19}.

Usually, starting from the expressions for the RG functions, one can
either develop the $\varepsilon$-expansion, or work directly at
$d=3$ considering the renormalized couplings $(g_1,g_2)$ as the expansion
parameters \cite{Schloms1,Schloms2}.
However, such RG perturbation theory series
are known to be asymptotic at best \cite{ZinnJustin02,Kleinert01}.
Moreover, equations for the FP are degenerated on a one-loop level in our particular case (see $\beta$-functions (\ref{b1}) and (\ref{b2})). For this case, $\sqrt\varepsilon$-expansion was elaborated \cite{Khmelnitskii75,Lubensky75,Grinstein76}, but it appears to have much worse convergent properties.
Therefore we rely on the analysis of Eqs. (\ref{b1})--(\ref{gam2}) at fixed $d=3$, putting there $\varepsilon=1$ and analysing them as functions
of couplings $g_1$ and $g_2$.

One should apply appropriate resummation techniques to improve the convergence of the RG perturbation series  to get reliable estimates.
Many different resummation procedures are currently 
used to this end, see e.g., Refs. \cite{Folk03,Pelissetto02,Holovatch02}.
In our study, we have used the method, based on the Borel
transformation combined with a conformal mapping \cite{LeGuillou77}.  
This resummation
technique was first elaborated for field-theoretical models with one coupling. It
requires knowledge about a high-order behavior of the series. 
This procedure was extended  for  field-theoretical
models that contain several couplings \cite{Alvarez00,Pelissetto00,Pelissetto04}. For instance, it was used for the study of frustrated magnets modeled by field-theoretical effective Hamiltonian with two couplings \cite{Pelissetto04,Delamotte08,Dudka10,Delamotte10}. 
Although Borel summability has not been proven for perturbative series  for models with
disorder in $d>0$ \cite{Alvarez00}, the conformal Borel method has been successfully used
for random Ising model \cite{Pelissetto00,Calabrese04,Folk06,Krinitsyn06}.
The following is a brief description of the procedure we use.

For the single variable case, the starting point is the  infinite
	series
	\begin{equation}
		f(g)=\sum^{\infty}_{n=0} a_n \ g^n  \label{series1}\,,
	\end{equation}
	with the coefficients $a_n$ growing factorially. Then,  one defines
	Borel-Leroy transform by:
	\begin{equation}
		B(u)=\sum^{\infty}_{n=0} \frac{a_n}{\Gamma[n+b+1]}  g^n,  \label{borelsum}
	\end{equation}
	where $\Gamma[\dots]$ is the Gamma-function. This transform is supposed to
	converge, in the complex plane, on the disk of radius $1/a$
	where $g=-1/a$ is the singularity of $B(g)$ nearest to the  origin.
	Then, using the integral
	representation of $\Gamma[n+b+1]$, $f(g)$ is rewritten as:
	\begin{equation}
		f(g)= \sum^{\infty}_{n=0} {\frac{a_n}{\Gamma[n+b+1]}}   g^n \int_0^{\infty} \
		d t \,\,{\rm e}^{-t} t^{n+b} .
	\end{equation}
	Then, interchanging summation and integration, one can define
	the Borel transform of $f$ as:
	\begin{equation}
		f_B(u)=\int_0^{\infty} d t\,\, {\rm e}^{-t} t^{b} B(gt).
		\label{boreltrans}
	\end{equation}

	In order  to compute the integral in  (\ref{boreltrans})  on the
	whole real positive  semi-axis one needs  an analytic continuation
	of $B(t)$.   It may be achieved by several methods. 

    Notice that by using the conformal mapping technique, we assume that  all the
	singularities of $B(u)$ lie on the negative real axis and that $B(u)$
	is analytic  in the whole complex plane excluding
	the cut from $-1/a$ to $-\infty$. Then,  one can perform in $B$
	the change of variable $w(g)=
        \frac{\sqrt{1 + a\, g}-1}{\sqrt{1 + a\, g}+1}
	\Longleftrightarrow      g(w)=\frac{4}{a}\frac{w}{(1-w)^2}$.
	This change maps the  complex $g$-plane cut from  $g=-1/a$  to $-\infty$
	onto the unit circle in the $w$-plane such that the singularities
	of $B(g)$ lying  on the negative axis  now lie on   the boundary
	of the  circle $|w|=1$. 
    
	Finally, the resulting expression of $B(w(g))$ has to be re-expanded in  powers of $w(g)$ and  the resumed expression of  the series $f$ writes:
	\begin{equation}
		f_{\mathrm{R}}(g)=\sum^{\infty}_{n=0} d_n(a, b) \int_0^{\infty}
		d t \,\,{{\rm e}^{-t} t^{b} \left[w(g t)\right]^n}, \label{resummation1}
	\end{equation}
	where     $d_n(a,b)$   are the coefficients of the re-expansion of
	$B(g(w))$ in the powers of $w(g)$.
	
	It is also useful to generalize the above expression (\ref{resummation1}) in the
	following way  \cite{kazakov79}
	\begin{equation}
		f_{\mathrm{R}}(g){=}\sum^{\infty}_{n=0} d_n(\alpha,a, b)  \int_0^{\infty}
	\!\!	d t\,\,{{\rm e}^{-t}\,  t^{b}}\ \frac{ \left[w(u
			t)\right]^n}{ \left[1-w(u t)\right]^{\alpha}}
		\label{resummation2}
	\end{equation}
	since this allows to impose the strong coupling behavior of the series: $f(g\to \infty)\sim g^{\alpha/2}$.
	
	This procedure can be extended  for  field-theoretical
	models that contain several couplings. For instance, when $f$ is a function  of two
	variables  $g_1$ and $g_2$,  the  resummation  technique can treat $f$ as a
	function of coupling associated with terms related to the pure (e.g., not disordered) magnet (in our case it is $g_2$) and
	a ratio $z=g_1/g_2$, \cite{Alvarez00,Pelissetto00,Pelissetto04}:
	\begin{equation}
		f(g_2,z)=\!\sum_{n{=}0}^\infty a_n(z) \ g_2^n \, . \label{series}
	\end{equation}
	Then, keeping $z$ fixed and performing the
	resummation only in $g_2$   according to the steps described above, one gets:
	\begin{equation}
		f_{\mathrm{R}}(g_2,z){=}\sum_{n=0}^\infty d_n(\alpha,a(z),b;z)\!
		\int_0^{\infty} \!\!d t\, {{\rm e}^{-t} t^{b}}\frac{
			\left[w(g_2 t;z)\right]^n}{ \left[1{-}w(g_2
			t;z)\right]^{\alpha} }  , \label{resummation}
	\end{equation}
	with $z$-dependent parameter $a(z)$. Here,  as     above,  the  coefficients
	$d_l(\alpha,a(z),b,z)$ in (\ref{resummation})  are computed  so
	that   the  re-expansion of  the right hand side  of (\ref{resummation}) in
	powers of $g_2$ coincides with that of (\ref{series}).
	
Calculation of  the parameters $a$, $b$ and  $\alpha$ for the field-theoretical models with several couplings
is a highly nontrivial task. Furthermore, in practice,  procedure described above is applied to the
truncated series (\ref{b1})-(\ref{gam2}), which are known up to the corresponding order of couplings.  As a result,
the resummed expansions (that usually correspond to certain physical observable) appear to be dependent
on resummation parameters $a$, $b$ and $\alpha$.
Often applying the conformal Borel method for models with two couplings, such
parameters  are taken in the region where physical observables are less sensitive to their values,
see e.g., Ref. \cite{Delamotte08}.
In our analysis we have tested the region of parameters discussed  in the study of five-loop series for the model
of frustrated magnet  \cite{Delamotte08}: $a=1/2$, $b$ is
varying in the interval  $[6, 30]$, while $\alpha$  is varying in $[-0.5, 2]$. As a final choice of the resummation parameters
we get $a=1/2$, $b=10$ and $\alpha=1$. At such values, the asymptotic
critical exponents $\nu$ and $\eta$ of the random Ising model are very close to the latest analytical estimates, see Table~\ref{table:data}.

\section{Quotients method}
\label{quotient}
In this appendix, we briefly describe the quotients method used to obtain extrapolated results for the critical exponents and scaling corrections presented in the paper.

Let $O(\beta,L)$ be a
dimensionful quantity that scales in the thermodynamic limit as
$\xi^{x_O/\nu}$. Notice that $x_O=0$ for dimensionless observables. Furthermore, we will denote 
by the symbol $g$ all the
dimensionless quantities, like $R_\xi$, $U_4$ or $g_2$.

The basis of the quotients method is to compare the same observable computed at $L$ and $2L$
\begin{equation}
{\mathcal
  Q}_O=\frac{O(\beta_\text{cross}(L, 2L),2L)}{O(\beta_\text{cross}(L, 2L),L)}\,,
  \end{equation}
  at $\beta_\text{cross}(L, 2L)$ which is defined by
  \begin{equation}
g(L,\beta_\text{cross}(L, 2L))=g(2L, \beta_\text{cross}(L, 2L))\,.
  \end{equation}

Therefore, one can write
\begin{equation}\label{eq:QO}
{\mathcal Q}_O^{\,\mathrm{cross}}=2^{x_O/\nu}+\mathcal{O}(L^{-\omega})\,,\
\end{equation}
and
\begin{equation}\label{eq:QOB}
g^{\,\mathrm{cross}}=g^* +\mathcal{O}(L^{-\omega})\,,\
\end{equation}
where $x_O/\nu$, $g^*$ and the correction-to-scaling exponent $\omega$
are universal quantities. To compute the $\nu$ and $\eta$
exponents, one should study dimensionful observables such as the
susceptibility ($x_\chi= \nu(2-\eta)$) and the $\beta$-derivatives of
$R_\xi$ and $U_4$, both having $x=1$.

The crossing point of the inverse
temperature ($\beta_\text{cross}(L,2L)$) scales as
\begin{equation}\label{eq:Tc}
\beta_\text{cross}(L,2L) = \beta_\text{c} + A_{\beta_\text{c},g} L^{-\omega -                                                                             
  1/\nu}+\ldots\,.
\end{equation}

Finally, the leading correction-to-scaling exponent can be estimated using the scaling of 
a dimensionless quantity $g$ ($Q_g$) via
\begin{equation}\label{eq:dimensionless-quotients}
\mathcal Q^{\text{cross}}_g(L) = 1 + A_g L^{-\omega} + B_g L^{-2\omega} + \ldots .
\end{equation}


%

\end{document}